\documentclass[12pt]{article}
\usepackage{amsmath}
\input amssym

\begin{document}

\def\prg#1{\medskip\noindent{\bf #1}}  \def\ra{\rightarrow}
\def\lra{\leftrightarrow}              \def\Ra{\Rightarrow}
\def\nin{\noindent}                    \def\pd{\partial}
\def\dis{\displaystyle}                \def\inn{\,\rfloor\,}
\def\grl{{GR$_\Lambda$}}               \def\Lra{{\Leftrightarrow}}
\def\cs{{\scriptstyle\rm CS}}          \def\ads3{{\rm AdS$_3$}}
\def\Leff{\hbox{$\mit\L_{\hspace{.6pt}\rm eff}\,$}}
\def\bull{\raise.25ex\hbox{\vrule height.8ex width.8ex}}
\def\ric{{(Ric)}}                      \def\tric{{(\widetilde{Ric})}}
\def\tmgl{\hbox{TMG$_\Lambda$}}
\def\Lie{{\cal L}\hspace{-.7em}\raise.25ex\hbox{--}\hspace{.2em}}
\def\sS{\hspace{2pt}S\hspace{-0.83em}\diagup}   \def\hd{{^\star}}
\def\dis{\displaystyle}                 \def\ul#1{\underline{#1}}

\def\hook{\hbox{\vrule height0pt width4pt depth0.3pt
\vrule height7pt width0.3pt depth0.3pt
\vrule height0pt width2pt depth0pt}\hspace{0.8pt}}
\def\semidirect{\;{\rlap{$\supset$}\times}\;}
\def\first{\rm (1ST)}       \def\second{\hspace{-1cm}\rm (2ND)}
\def\bm#1{\hbox{{\boldmath $#1$}}}
\def\nb#1{\marginpar{{\large\bf #1}}}

\def\G{\Gamma}        \def\S{\Sigma}        \def\L{{\mit\Lambda}}
\def\D{\Delta}        \def\Th{\Theta}
\def\a{\alpha}        \def\b{\beta}         \def\g{\gamma}
\def\d{\delta}        \def\m{\mu}           \def\n{\nu}
\def\th{\theta}       \def\k{\kappa}        \def\l{\lambda}
\def\vphi{\varphi}    \def\ve{\varepsilon}  \def\p{\pi}
\def\r{\rho}          \def\Om{\Omega}       \def\om{\omega}
\def\s{\sigma}        \def\t{\tau}          \def\eps{\epsilon}
\def\nab{\nabla}      \def\btz{{\rm BTZ}}   \def\heps{\hat\eps}
\def\bu{{\bar u}}     \def\bv{{\bar v}}     \def\bs{{\bar s}}
\def\bx{{\bar x}}     \def\by{{\bar y}}
\def\tphi{{\tilde\vphi}}  \def\tt{{\tilde t}}

\def\tG{{\tilde G}}   \def\cF{{\cal F}}      \def\bH{{\bar H}}
\def\cL{{\cal L}}     \def\cM{{\cal M }}     \def\cE{{\cal E}}
\def\cH{{\cal H}}     \def\hcH{\hat{\cH}}
\def\cK{{\cal K}}     \def\hcK{\hat{\cK}}    \def\cT{{\cal T}}
\def\cO{{\cal O}}     \def\hcO{\hat{\cal O}} \def\cV{{\cal V}}
\def\tom{{\tilde\omega}}                     \def\cE{{\cal E}}
\def\cR{{\cal R}}    \def\hR{{\hat R}{}}     \def\hL{{\hat\L}}
\def\tb{{\tilde b}}  \def\tA{{\tilde A}}     \def\tv{{\tilde v}}
\def\tT{{\tilde T}}  \def\tR{{\tilde R}}     \def\tcL{{\tilde\cL}}
\def\hy{{\hat y}\hspace{1pt}}  \def\tcO{{\tilde\cO}}

\def\nn{\nonumber}                    \def\vsm{\vspace{-1pt}}
\def\be{\begin{equation}}             \def\ee{\end{equation}}
\def\ba#1{\begin{array}{#1}}          \def\ea{\end{array}}
\def\bea{\begin{eqnarray} }           \def\eea{\end{eqnarray} }
\def\beann{\begin{eqnarray*} }        \def\eeann{\end{eqnarray*} }
\def\beal{\begin{eqalign}}            \def\eeal{\end{eqalign}}
\def\lab#1{\label{eq:#1}}             \def\eq#1{(\ref{eq:#1})}
\def\bsubeq{\begin{subequations}}     \def\esubeq{\end{subequations}}
\def\bitem{\begin{itemize}}           \def\eitem{\end{itemize}}
\renewcommand{\theequation}{\thesection.\arabic{equation}}

\title{Siklos waves with torsion in 3D}

\author{M. Blagojevi\'c and B. Cvetkovi\'c\footnote{
        Email addresses: {mb@ipb.ac.rs,
                          cbranislav@ipb.ac.rs}}\\
Institute of Physics, University of Belgrade \\
                      Pregrevica 118, 11080 Belgrade--Zemun, Serbia}
\date{\today}
\maketitle

\begin{abstract}
Starting from the Siklos waves in general relativity with a cosmological
constant, interpreted as gravitational waves on the anti-de Sitter
background, a new class of exact torsion waves is constructed in the
framework of three-dimensional gravity with propagating torsion. In the
asymptotic limit, the geometry of torsion waves takes the anti-de Sitter
form. In the sector with massless torsion modes, we found a set of
asymptotic conditions that leads to the conformal asymptotic symmetry.
\end{abstract}

\section{Introduction}
\setcounter{equation}{0}

Exact gravitational waves have been an important subject of investigation
in general relativity (GR) from the early 1920s; for a review, see
\cite{1,2,3,4}. Most of the activity on the subject has been focused on
asymptotically flat models, the solutions of GR without a cosmological
constant. From 1980s, exact gravitational waves have been studied also in
GR with a cosmological constant (\grl) \cite{5,6,7}, see also \cite{8};
for higher-dimensional extensions, see \cite{9}. In particular, exact
gravitational waves with an AdS asymptotic behavior attracted a lot of
interest in regard to the AdS/CFT correspondence \cite{10}. Moreover, some
of these solutions ``may serve as exact models of the propagation of
primordial gravitational waves and may be relevant for the (hypothetical)
cosmological wave background" \cite{11}.

To properly understand dynamical complexities of gravity, one often relies
on technically simplified three-dimensional (3D) models \cite{12}. In 3D,
both GR and \grl\ are topological theories without propagating degrees of
freedom, in which nontrivial wave solutions can exist only in the presence
of \emph{matter sources} \cite{13,14}. To avoid such a degenerate
situation, one is naturally motivated to study alternative gravitational
models possessing true dynamical degrees of freedom. The well-known models
of this type, formulated in the context of \emph{Riemannian geometry} of
spacetime, are Topological massive gravity and New massive gravity
\cite{15,16}. Their dynamical properties allow for the existence of
gravitational waves \emph{in vacuum}; see, for instance, Ay\'on-Beato et
al. \cite{17}.

In the early 1960s, a new approach to gravitational dynamics was proposed,
based on a modern, gauge-field-theoretic approach, known as the
\emph{Poincar\'e gauge theory} (PGT) \cite{18,19,20}, with an underlying
\emph{Riemann--Cartan (RC) geometry} of spacetime, characterized by both
the \emph{curvature} and the \emph{torsion}. In a topological version of
the three-dimensional PGT, gravitational waves with torsion were
constructed in the presence of matter sources by Obukhov \cite{21}.
However, genuine gravitational waves are those that can propagate in
spacetime regions without matter. Further investigations of the PGT, with
a Lagrangian that is at most quadratic in the field strengths (quadratic
PGT), revealed a rich dynamical structure, expressed, in particular, by
the existence of propagating torsion modes \cite{22}. In a recent paper
\cite{23}, we used quadratic PGT to construct exact torsion waves in
vacuum as a generalization of the plane-fronted waves from GR.

In the present paper, we continue the investigation of genuine
gravitational waves with torsion in 3D, by focusing on the anti-de Sitter
(AdS) background. We found a new class of exact torsion waves in vacuum,
representing a PGT extension of the Siklos waves in \grl\ \cite{24}, see
also \cite{8,11}. In the linear approximation, this class is associated to
spin-$2$ torsion excitations around the AdS background. In the sector of
massless torsion modes, we found a set of asymptotic conditions that leads
to a conformal asymptotic symmetry, characterized by two independent
Virasoro algebras with central charges. On the other hand, massive torsion
waves show kind of an oscillatory behavior in the asymptotic region.

The paper is organized as follows. In section 2, we give an overview of
the Siklos waves in the three-dimensional \grl. In section 3, we construct
a new wave solution in PGT, taking the metric to be of the Siklos form,
whereas the torsion piece of the connection is assumed to possess only the
tensorial irreducible component. The solutions of the field equations are
found and classified according to the values of the mass parameter $\m^2$,
associated to the spin-$2$ torsion modes. For $\m^2\ge 0$ (no tachyons),
the asymptotic limit of the Siklos waves with torsion is shown to be
represented by Riemannian AdS spacetimes. In section 4, we study the form
of the AdS asymptotic conditions for $\m^2\ge 0$. It turns out that a
well-defined asymptotic structure exists only in the massless sector. The
corresponding central charges of the asymptotic symmetry are found in
section 5,  and section 6 is devoted to concluding remarks. Finally, three
appendices contain some technical details.

Here are our conventions: the Latin indices $(i, j, k, ...)$ refer
to the local Lorentz (co)frame and run over $(+,-,2)$,  $b^i$ is the
triad field (coframe 1-form), $h_i$ is the dual basis (frame), totally
antisymmetric tensor $\ve^{ijk}$ is normalized to $\ve^{+-2}=1$; the Greek
indices $(\m,\n,\r, ...)$ refer to the coordinate frame; the Lie dual of
an antisymmetric form $X^{jk}$ is $X_i:=-\ve_{ijk}X^{jk}/2$, the Hodge
dual of a form $\a$ is $\hd\a$, and the exterior product of forms is
implicit.

\section{Siklos waves}
\setcounter{equation}{0}

In 1980s, Siklos \cite{24} found a special class of exact gravitational
waves propagating on the AdS background, the physical interpretation of
which was investigated in detail by Podolsk\'y \cite{11}. In the
Poincar\'e coordinates $x^\m=(u,v,y)$, the Siklos metric in 3D has the
form
\be
ds^2=\frac{\ell^2}{y^2}\left[2du(Hdu+dv)-dy^2\right]\, ,        \lab{2.1}
\ee
with $H=H(u,y)$, which is equivalent to a subclass of the Kundt metric
\cite{8,11}. The wave fronts are labeled by $u=$ const.,  $v$ is an affine
parameter along the corresponding rays generated by the Killing vector
field $\pd_v$ that is null but not covariantly constant,  and for $H=0$
the metric reduces to the AdS background (see Appendix A). We choose the
triad field $b^i$ (1-form) to be
\be
b^+:=\frac{\ell}{y}du\, ,\quad
b^-:=\frac{\ell}{y}(Hdu+dv)\, ,\quad
b^2=\frac{\ell}{y}dy\, ,                                        \lab{2.2}
\ee
so that the line element is given by $ds^2=\eta_{ij} b^ib^j$, with the
half-null Lorentz metric
$$
\eta_{ij}=\left( \ba{ccc}
             0 & 1  & 0  \\
             1 & 0  & 0  \\
             0 & 0  & -1
                \ea
       \right)\, .
$$
The dual frame basis $h_i$, defined by $h_i\hook b^j=\d_i^j$, is given by
$$
h_+=\frac{y}{\ell}(\pd_u-H\pd_v)\,,\qquad
h_-=\frac{y}{\ell}\pd_v\, ,\qquad h_2=\frac{y}{\ell}\pd_y\,.
$$

The related Riemannian connection $\om^{ij}$ (1-form) can be written in a
compact form as
\bsubeq\lab{2.3}
\be
\om^{ij}=
  \bar\om^{ij}-\frac{1}{\ell}\ve^{ij}{_m}k^m(yH')k_nb^n\, .     \lab{2.3a}
\ee
Here, prime denotes a derivative with respect to $y$, the first term
$\bar\om^{ij}$ describes the background AdS geometry,
\be
\bar\om^{+-}=0\, ,\qquad
    \bar\om^{+2}=\frac{1}{\ell}b^+\,,
       \qquad \bar\om^{-2}=\frac{1}{\ell}b^-\, ,                \lab{2.3b}
\ee
\esubeq
and the second one is the radiation piece, characterized by the null
vector $k^m:=(0,1,0)$, with $k_m=(1,0,0)$.

Next, we calculate the Riemannian curvature,
\bsubeq
\be
R^{ij}=\frac{1}{\ell^2}b^ib^j
       -\frac{1}{\ell^2}\ve^{ij}{_m}k^m(y^2H''-yH')k^n\hd b_n\, ,
\ee
whereupon the Ricci curvature $\ric^i=-h_j\hook R^{ij}$ and the scalar
curvature $R=h_i\hook\ric^i$ are found to be
\bea
&&\ric^i=\frac{2}{\ell^2}b^i+\frac{1}{\ell^2}k^i(y^2H''-yH')k_nb^n\, ,\nn\\
&&R=\frac{6}{\ell^2}\, .
\eea
\esubeq

When the Siklos metric satisfies the vacuum field equation of \grl\ with
$\L\sim -1/\ell^2$, the metric function $H$ takes a simple form:
\be
y^2H''-yH'=0 \quad\Ra\quad H=D_1(u)+D_2(u)y^2\, .
\ee
However, this solution is \emph{trivial}. Indeed, since the radiation
piece of the curvature vanishes on shell, we have $R^{ij}=b^ib^j/\ell^2$,
and the geometry of spacetime is fixed, it has the AdS form. Nontrivial
AdS waves can exist in \grl\ only in the presence of matter \cite{13,14},
but to have \emph{vacuum} AdS waves, one has to change the gravitational
dynamics. As we shall see, transition to quadratic PGT allows the
existence of genuine AdS waves with torsion.

\section{Siklos waves with torsion}
\setcounter{equation}{0}

Basic gravitational variables of PGT are the triad field $b^i$ and the
Lorentz connection $\om^{ij}$ (1-forms), and the related field strengths
are the torsion $T^i=db^i+\om^i{_m}b^m$ and the curvature
$R^{ij}=d\om^{ij}+\om^i{_m}\om^{mj}$ (2-forms). Relying on PGT, we now
introduce a geometric extension of the Siklos waves \eq{2.1} to genuine
Siklos waves with torsion.

\subsection{Ansatz}

In order to \emph{preserve the radiation nature} of the Siklos metric, we
assume that the form of the triad field in PGT remains the same as in Eq.
\eq{2.2}. Essentially the same idea can be applied also to the connection
\cite{23}: starting from the Riemannian connection \eq{2.3}, we assume
that the new, RC connection is given by
\bsubeq\lab{3.1}
\be
\om^{ij}=\bar\om^{ij}-\frac{1}{\ell}\ve^{ij}{_m}k^m(yG)k_nb^n\,,
\ee
where
\be
G:=H'+K\, ,\qquad K=K(u,y)\, .
\ee
\esubeq
Geometrically, the new function $K$ in the connection is related to the
torsion:
\be
T^i:=\nab b^i=-\frac{yK}{\ell}k^ik^n\hd b_n\, .
\ee
For $K=0$, the torsion vanishes, and the connection becomes equivalent to
$\bar\om^{ij}$. The only nonvanishing irreducible component of $T^i$ is
its tensorial piece ${}^{(1)}T^i$ \cite{23}, so that
$$
{}^{(1)}T^i=T^i\, .
$$

Using the above ansatz for the connection, one can calculate the RC
curvatures:
\bea
&&R^{ij}=\frac{1}{\ell^2}b^ib^j
   - \frac{1}{\ell^2}\ve^{ij}{_m}k^m (y^2G'-yH')k^n\hd b_n\, ,  \lab{3.3}\\
&&\ric^i=\frac{2}{\ell^2}b^i+\frac{1}{\ell^2}k^i(y^2G'-yH')k_n b^n\,,\nn\\
&&R=\frac{6}{\ell^2}\, .                                        \nn
\eea
The quadratic curvature invariant takes the form
$$
R^{ij}\hd R_{ij}=\frac{6}{\ell^4}\hd 1\, .
$$
The only nonvanishing irreducible components of $R^{ij}$ are:
$$
{}^{(6)}R^{ij}=\frac{1}{6}Rb^ib^j\, ,\qquad
{}^{(4)}R^{ij}=R^{ij}-{}^{(6)}R^{ij}\,.
$$
For more details on the irreducible decomposition of the field strengths,
see Ref. \cite{23}.

In what follows, the specific forms of both the metric function $H$ and
the torsion function $K$ will be determined by the PGT field equations.

\subsection{Lagrangian dynamics of PGT}

The PGT dynamics is described by a Lagrangian 3-form
$L_G=L_G(b^i,T^i,R^{ij})$, which is assumed to be at most quadratic in the
field strengths (quadratic PGT) and parity invariant. In conformity with
our ansatz, the Lagrangian is chosen to have the form
\bea
L_G&=&-a_0\ve_{ijk}b^i R^{jk}
      -\frac{1}{3}\L_0\ve_{ijk}b^i b^j b^k                      \nn\\
   &&+T^i\,\hd(a_1{}^{(1)}T_i)
     +\frac{1}{2}R^{ij}\,\hd(b_4{}^{(4)}R^{ij}+b_6{}^{(6)}R^{ij})\,.\lab{3.4}
\eea
Indeed, the only nonvanishing irreducible components of the field
strengths appearing in $L_G$ are ${}^{(1)}T^i$,  ${}^{(4)}R^{ij}$ and
${}^{(6)}R^{ij}$, and $a_1,b_4,b_6$ are the corresponding coupling
constants. Then, the PGT field equations in vacuum are found to be (see
Appendix B):
\bea
&&(1ST):\quad
  (a_0\ell^2-b_4-b_6)(yH''-H')+(a_0\ell^2-a_1\ell^2-b_4-b_6)yK'=0\,,\nn\\
&&\hspace{2cm} 2a_0\ell^2+b_6+2\ell^4\L_0=0\, ,                 \nn\\
&&(2ND):\quad
  b_4\left[y^2(H'''+K'')+yK'\right]-(a_0\ell^2-a_1\ell^2-b_6)K=0\,.\lab{3.5}
\eea
These equations are checked using the Excalc package of the computer
algebra system Reduce. Using the expression for $(1ST)'$, one finds that
$(2ND)$ can be rewritten as
$$
y^2K''+yK'+\ell^2\m^2 K=0\, ,\qquad
   \m^2=\frac{(a_1-a_0-b_6\l)(a_0+b_4\l+b_6\l)}{b_4a_1}\, ,
$$
where $\l:=-1/\ell^2$. Finally, after introducing the notation
$$
\hy=\frac{y}{\ell}\, ,\qquad m^2=\ell^2\m^2\,,
$$
the two field equations take a more compact form:
\bea
&&(1ST)\qquad \hy H''-H'=\ell C\hy K'\, ,\qquad
          C:=\frac{a_1}{a_0+b_4\l+b_6\l}-1\, ,                  \nn\\
&&(2ND)\qquad \hy^2K''+\hy K'+m^2 K=0\, ,                       \lab{3.6}
\eea
where prime now denotes differentiation with respect to $\hy$. As one can
see, it is the presence of torsion ($K\ne 0$) that makes the metric of the
AdS wave nontrivial ($\hy H''-H'\ne 0$). Equations \eq{3.6} define a new
class of Siklos waves---the Siklos waves with torsion.

\subsection{Solutions}

The coefficient $m^2$ in $(2ND)$ is the (dimensionless) mass parameter
associated to the spin-2 excitation of the torsion field around the AdS
background, see \cite{22,23}. The absence of tachyons requires $m^2\ge 0$.
In this subsection, we construct the exact Sikos waves with torsion, and
classify them according to the values of $m^2$.

\prg{\bm{(1)~ m^2>0.}}
The Euler (or Euler-Fuchs, Euler-Cauchy) differential equation $(2ND)$  is
solved by the ansatz $K=\hy^\a$, which yields $\a^2+m^2=0$. For $m^2>0$, we
have $\a=\pm im$, so that $K=\hy^{\pm im}=e^{\pm im\ln\hy}$, or equivalently,
\bsubeq\lab{3.7}
\be
K=A(u)\cos(m\ln\hy)+B(u)\sin(m\ln\hy)\, .
\ee
By substituting this result into $(1ST)$, one finds the related solution
for $H$:
\be
H=D_1+D_2\hy^2+\frac{\ell Cm}{1+m^2}
               \hy\left[A(u)\sin(m\ln\hy)-B(u)\cos(m\ln\hy)\right]\, .
\ee
\esubeq
The first two terms, which represent a solution of the homogeneous
equation $\hy H''-H'=0$, can be geometrically disregarded, as they do not
influence the values of the field strengths.

In the asymptotic limit $\hy\to 0$, the torsion and the radiation piece of
the curvature, ${}^{(4)}R_{ij}$, vanish, as follows from the relations
\bea
&&\lim_{\hy\to 0}\hy K=0\, ,                                        \nn\\
&&\lim_{\hy\to 0}\Bigl[\hy^2(H''+K')-\hy H'\Bigr]
  =\lim_{\hy\to 0}\Bigl[\hy^2K'+\ell C\hy^2K'\Bigr]=0\, .
\eea
Thus, the asymptotic geometry of our solution is given by the Riemannian
AdS spacetime.

\prg{\bm{(2)~ m^2=0.}}
In order to have a smooth Minkowskian limit for $\ell^2\to\infty$, the
condition $m^2=0$ is realized by demanding \cite{22}
\be
a_1-a_0+b_6/\ell^2=0\, .                                        \lab{3.9}
\ee
As a consequence, the solution for the massless torsion wave is given by
\bea
&& K=C_1+C_2\ln\hy\, ,                                          \nn\\
&& H=D_1+D_2\hy^2-\ell C C_2\hy\, .                             \lab{3.10}
\eea
As before, one can choose $D_1=D_2=0$ without loss of generality, so that
the asymptotic limit of the solution is again given by the Riemannian AdS
spacetime.

\prg{\bm{(3)~ m^2<0.}}
Although the spin-2 torsion modes are now tachyons, we present the related
exact wave solution, for the sake of completeness:
\bea
&&K=A\hy^{|m|}+B\hy^{-|m|}\, ,                                        \nn\\
&&H=\frac{\ell C|m|}{|m|^2-1}\left(A\hy^{1+|m|}-B\hy^{1-|m|}\right)\, .
\eea
The asymptotic behavior depends on the value of $|m|$.

\section{Asymptotic conditions}
\setcounter{equation}{0}

In our study of the asymptotic conditions, we assume that the topology of
the spacetime manifold $M$ is $R\times\S$, where $R$ is interpreted as
time, and $\S$ is a spatial section of spacetime, whose boundary $\pd\S$
is topologically a circle. The asymptotic analysis is simplified by
introducing a new set of local coordinates $(t,\vphi)$, given by
$u=(t+\ell\vphi)/\sqrt{2}$, $v=(t-\ell\vphi)/\sqrt{2}$, such that the
boundary $\pd\S$ at $y=0$ is parametrized by the angular coordinate
$\vphi$.

As we have seen in the previous section, in the asymptotic limit $y\to 0$,
the geometry of our torsion wave is described by the Riemannian AdS
spacetime. This property motivates us to examine asymptotic conditions
based on the following requirements:
\bitem
\item[(a)] asymptotic configurations include the torsion wave
    geometry;\vsm
\item[(b)] they are invariant under the action of the
    AdS group $SO(2,2)$;\vsm
\item[(c)] asymptotic symmetries have well defined canonical
generators.
\eitem
Specific aspects of these criteria depend on the value of the mass
parameter $\m^2$.

\subsection{Massive torsion waves}

For $\m^2>0$, the characteristic functions $H$ and $K$ can be represented
in the form
\bsubeq\lab{4.1}
\be
H=yW_0\, ,\qquad  K=W_0\, ,
\ee
where $W_0$ is a \emph{generic} wave ``oscillatory" function,
\be
W_0:=C_1(u)\cos(m\ln y/\ell) + C_2(u)\sin(m\ln y/\ell)\, .
\ee
\esubeq
In spite of this oscillatory behavior, both the torsion and the wave piece
of the curvature tend to zero when $y\to 0$.

In the matrix notation, the components of the Siklos metric \eq{2.1} read
$$
g_{\m\n}=\frac{\ell^2}{y^2}
           \left( \ba{ccc}
            2H & 1  & 0  \\
             1 & 0  & 0  \\
             0 & 0  & -1
                \ea
         \right)\, .
$$
Asymptotically, for $y\to 0$, we have $g_{uu}\sim W_0/y$, so that, to
leading order in $1/y$, $g_{\m\n}$ reduces to the AdS metric $\bar
g_{\m\n}$. In the asymptotic analysis, we use $\cO(y^nW_0)$ to denote a
term that is \emph{at most} proportional to $y^nW_0$ when $y\to 0$. Thus,
the Siklos metric is of the type
$$
g_{\m\n}=\bar g_{\m\n}+G_{\m\n}\, ,\qquad
              G_{\m\n}:=\left( \ba{ccc}
                                \cO(W_0/y) & 0 & 0  \\
                                 0         & 0 & 0  \\
                                 0         & 0 & 0
                               \ea\right) \, .
$$
Looking at the action of the AdS Killing vectors (Appendix A) on
$g_{\m\n}$, one finds that the general requirements (a) and (b) are
fulfilled by the following asymptotic configurations:
\be
g_{\m\n}=\bar g_{\m\n}+G_{\m\n}\,,\qquad
G_{\m\n}:=\left( \ba{ccc}
                   \cO_{-1} & \cO_0  & \cO_0  \\
                   \cO_0    & \cO_0  & \cO_0  \\
                   \cO_0    & \cO_0  & \cO_0
                 \ea\right) \, ,                                \lab{4.2}
\ee
where $\cO_n:=\cO(y^nW_0)$. The asymptotic form \eq{4.2}, but with
$\cO_n=\cO(y^n)$, was studied earlier by Afshar et al. \cite{25,26}, in
the context of Conformal Chern--Simons gravity.

The asymptotic conditions \eq{4.2} are preserved by the local translations
of the form
\bea
&&\xi^u=\ve^u(u)+\frac{y^2}{4}\pd_v^2\ve^v(v)+\cO_3\, ,         \nn\\
&&\xi^v=\ve^v(v)+\frac{y^2}{4}\pd_u^2\ve^u(u)+\cO_3\, ,         \nn\\
&&\xi^2=\frac{y}{2}(\pd_u\ve^u+\pd_v\ve^v)+\cO_3\, .            \lab{4.3}
\eea
These parameters are essentially of the Brown-Henneaux type \cite{26,27}.

In  the next step, one could try to extend these considerations to the
variables $b^i$ and $\om^{ij}$. However, a problem arises when we return
to our general requirement (c). Namely, although the field strengths $T^i$
and $R^{ij}$ have an AdS asymptotic limit, the asymptotic behavior of
$b^i$ and $\om^{ij}$ is determined by the function $W_0$, which oscillates
when $y\to 0$. Thus, the basic dynamical variables have \emph{no
asymptotic limit}, and one is not able to define surface terms of the
canonical generators. Thus, one cannot formulate a boundary theory, and in
particular, the AdS/CFT correspondence is not well defined.

\subsection{Massless torsion waves}

In the sector with massless torsion modes, the form of our wave solution
is displayed in Eq. \eq{3.10}. As we noted before, the geometrically
irrelevant term $D_1+D_2y^2$ in $H$ can be removed by choosing
$D_1=D_2=0$, whereupon the characteristic functions $H$ and $K$ are of the
generic form
\be
H=C_0(y/\ell)\, ,\qquad  K=C_1+C_2\ln(y/\ell)\, .               \lab{4.4}
\ee
The asymptotic geometry of the solution is described by the AdS spacetime.
In this section, we discuss the asymptotic structure of the massless
torsion wave \eq{4.4}.

Quite generally, the wave triad \eq{2.2} can be written in the form
$b^i{_\m}=\bar b^i{_\m}+B^i{_\m}$, where $\bar b^i$ is the AdS triad, and
the only nonvanishing component of $B^i{_\m}$ is $B^-{_u}=\ell H/y=C_0$. Then,
in accordance with the general requirements (a) and (b), we choose the
following asymptotic form of the triad field:
\be
b^i{_\m}=\bar b^i{_\m}+B^i{_\m}\,,\qquad
B^i{_\m}:=\left( \ba{lll}
         \cO_1  & \cO_1  & \cO_1  \\
         \cO_0  & \cO_1  & \cO_1  \\
         \cO_1  & \cO_1  & \cO_1
               \ea\right)\, ,                                    \lab{4.5}
\ee
where $\cO_n:=\cO(y^n)$. These conditions impose the following restriction
on the local Poincar\'e parameters $(\xi^\r,\ve^{ij})$:
$$
\d_0 b^i{_\m}
  :=\ve^{ijk}\th_j b_{k\m}-(\pd_\m\xi^\r)b^i{_\r}-\xi^\r\pd_\r b^i{_\m}
   =B^i{_\m}\,,
$$
where $\th^i$ is the Lie dual of $\ve_{mn}$. As a consequence, the
asymptotic parameters of local translations take the form displayed in Eq.
\eq{4.3}, whereas the asymptotic parameters of Lorentz rotations are found
to be
\bea
&&\th^+=\frac{y}{2}\pd_v^2\ve^v+\cO_2\, ,                   \nn\\
&&\th^-=-\frac{y}{2}\pd_u^2\ve^u+\cO_2\, ,                  \nn\\
&&\th^2=\frac{1}{2}(\pd_v\ve^v-\pd_u\ve^u)+\cO_2\, .            \lab{4.6}
\eea

Next, we wish to examine whether the asymptotic behavior of the RC
connection \eq{3.1} can be made compatible with the already found form of
the asymptotic Poincar\'e parameters. First, we introduce the Lie-dual
connection $\om^i$:
\be
\om^+=\frac{1}{\ell}b^+\, ,\qquad
\om^-=-\frac{1}{\ell}b^-+\frac{y}{\ell}Gb^+\, ,\qquad \om^2=0\,.\lab{4.7}
\ee
The form of $K$ implies that the asymptotic conditions on the connection
should contain log terms. By combining the expression \eq{4.7} for
$\om^i{_\m}$ with the asymptotic formulas for $b^\pm$ and $G=H'+K$, we
find it suitable to assume
\be
\om^i{_\m}=\bar\om^i{_\m}+\Om^i{_\m}\,,\qquad
\Om^i{_\m}:=\frac{1}{\ell}
           \left( \ba{lll}
   \cO_1            &  \cO_1 & \cO_1            \\
   \cO(\ln y/\ell)  &  \cO_1 & \cO(y\ln y/\ell) \\
   \cO(y\ln y/\ell) &  \cO_1 & \cO_1
           \ea\right)\, .                                       \lab{4.8}
\ee
As it turns out, the asymptotic invariance of $\om^i{_\m}$,
$$
\d_0\om^i{_\m}:=-\pd_\m\th^i-\ve^{ijk}\om_{j\m}\th_k
     -\pd_\m\xi^\r\om^i{_\r}-\xi^\r\pd_\r\om^i{_\m}=\Om^i{_\m}\,,
$$
does not impose any new restriction of the asymptotic Poincar\'e
parameters \eq{4.3} and \eq{4.6}.

In order to clarify the interpretation of our asymptotic conditions, we
wish to find the commutator algebra of the asymptotic Poincar\'e
transformations. To do that, we note that the composition law of the
asymptotic transformations, to lowest order in $y$, reads
\be
(\ve^u)'''=(\ve^u)'\pd_u(\ve^u)''-(\ve^u)''\pd_u(\ve^u)'\, ,    \lab{4.9}
\ee
and similarly for $\ve^v$. Then, introducing the notation
$$
\ell^+_n:=-\frac{1}{\sqrt{2}}
          \d_0(\ve^u=\ell e^{inu\sqrt{2}/\ell},\ve^v=0)\,,\qquad
\ell^-_n:=-\frac{1}{\sqrt{2}}
          \d_0(\ve^u=0,\ve^v=\ell e^{inv\sqrt{2}/\ell})\,,
$$
the commutator algebra of the asymptotic symmetry takes the form of two
independent Virasoro algebras:
\be
i[\ell^\pm_m,\ell^\pm_n]=(m-n)\ell^\pm_{m+n}\, .                \lab{4.10}
\ee
The related central charges are discussed in the next section.

\section{Canonical form of the asymptotic symmetry}\label{sec5}
\setcounter{equation}{0}

In this section, we use the canonical approach to analyze the asymptotic
symmetry in the massless sector, including the values of the central
charges.

To simplify the analysis, we follow Nester \cite{28} in applying the
\emph{first-order formulation} to the quadratic PGT. In this formalism,
the Lagrangian \eq{3.4} is written in the form
\be
L_G=T^i\t_i+\frac{1}{2}R^{ij}\r_{ij}-V(b^i,\t_i,\r_{ij})
                            -\frac{1}{3}\L\ve_{ijk}b^ib^jb^k\, .\lab{5.1}
\ee
Here, $\t_i$ and $\r_{ij}$ are new, \emph{independent} variables, and $V$
is a function quadratic in $\t_i$ and $\r_{ij}$, chosen so that, on shell,
we have $\t_i=H_i$ and $\r_{ij}=H_{ij}$, where $H_i=\pd L_G/\pd T^i$ and
$H_{ij}=\pd L_G/\pd R^{ij}$ are the covariant field momenta associated to
the original Lagrangian \eq{3.4}. Explicit form of $V$ is described in
Ref. \cite{22}, and it ensures the first-order formulation \eq{5.1} to be
\emph{equivalent} to \eq{3.4}. Thus, the variation of $L_G$ with respect
to $\t_i$ and $\r_{ij}$ yields
\bea
&&\t_i=2 a_1\hd T_i\, ,                                         \nn\\
&&\r_{ij}=-2\left(a_0-\frac{1}{6}b_6R\right)\ve_{ijk}b^k
          +2b_4\hd {}^{(4)}R_{ij}\, ,                           \lab{5.2}
\eea
in accordance with the forms of $H_i$ and $H_{ij}$ defined by  the
Lagrangian \eq{3.4}.

Asymptotic symmetries are best described in the canonical formalism. In
the first order formulation of PGT, the canonical gauge generator is a
functional $G[\vphi,\pi]$ on the phase space, the form of which is defined
in Eqs. (5.7) of Ref. \cite{22}. The canonical generator acts on the
phase-space variables $(\vphi,\pi)$ via the Poisson (or Dirac) bracket
operation, defined in terms of the functional derivatives. A functional
$F[\vphi,\pi]=\int d^2 x f(\vphi,\pd_\a\vphi,\pi,\pd_\a\pi)$ is
differentiable (or regular) if its variation has the form $\d F=\int
d^2x\left[A(x)\d\vphi+B(x)\d\pi\right]$.  In order to ensure this property
for our generator $G$, we have to improve its form by adding an
appropriate surface term $\G$ \cite{29}. The improved canonical generator
$\tG:=G+\G$ has been calculated in Appendix C; it is both finite and
differentiable (well-defined).

The Poisson bracket (PB) algebra of the improved generators could be found
by a direct calculation, but we rather rely on another, more instructive
method. Introducing a convenient notation, $\tG'=\tG[\ve^u{}',\ve^v{}']$
and similarly for $\tG''$ and $\tG'''$, we use the main theorem of Ref.
\cite{30}, which states that the PB of two well-defined generators must
also be a well-defined generator, to conclude that the PB algebra has the
form
\bsubeq\lab{5.3}
\be
\{\tG'',\tG'\}=\tG'''+C'''\, .                                  \lab{5.3a}
\ee
Here, the parameters of $\tG'''$ are defined by the composition law
\eq{4.9}, and $C'''$ is the central charge of the algebra. A simple
reformulation of this formula, given by
\bea
\{\tG'',\tG'\}=\d_0' \tG''\approx \d_0'\G''\, ,
\eea
represents a powerful tool for calculating the central charge. Indeed, the
previous two equations imply
\be
\d_0'\G''\approx \G'''+C'''\, .
\ee
\esubeq
Now, since $C'''$ does not depend on the basic dynamical variables and
$\G'''$ vanishes on the AdS background (see Appendix C), the evaluation of
$\d_0'\G''$ on the AdS background yields the final expression for $C'''$:
\be
\overline{\d_0'\G''}=C'''\, .
\ee
An explicit calculation based on the results of Appendix C yields
\be
\frac{\sqrt 2}{\ell}C'''=
  -\left(a_0-\frac{b_6}{\ell^2}\right)\int_0^{2\pi}d\vphi
   \left(\ve^u{}''\pd_u^3\ve^u{}'+\ve^v{}''\pd_v^3\ve^v{}'\right)\,.
\ee
This result, combined with Eq. \eq{5.3a}, completes the derivation of the
canonical PB algebra.

A more familiar form of this algebra is obtained by introducing the
Fourier modes of the improved generator:
$$
L_n^+:=-\frac{1}{\sqrt 2}\tG(\ve^u=\ell e^{inu\sqrt2/l},\ve^v=0)\,,\qquad
L_n^-:=-\frac{1}{\sqrt 2}\tG(\ve^v=\ell e^{inv\sqrt2/l},\ve^u=0)\, .
$$
Then, the canonical algebra \eq{5.3a} takes the form of two independent
Virasoro algebras with central charges,
\be
i\{L_m^\pm,L_n^\pm\}=(m-n)L^\pm_{m+n}+\frac{c^\pm}{12}m^3\d_{m+n}\,,
\ee
where the central charges are equal to each other:
\be
c^\pm=\left(1-\frac{b_6}{a_0\ell^2}\right)c_0\, .
\ee
Note that the coupling constant $b_6$ modifies the \grl\ central charge
$c_0:=3\ell/2G$, and for $b_6<a_0\ell^2$, the central charge $c^\pm$ is
positive.

\section{Concluding remarks}

In this paper, we found a new class of exact vacuum solutions of the
three-dimensional PGT, the class of Siklos waves with torsion. Asymptotic
geometry of these solutions is described by the Riemannian AdS spacetime.
In the sector of massless torsion modes, we found a set of asymptotic
conditions for which the asymptotic symmetry is described by two
independent Virasoro algebras with equal central charges $c^\pm$, the
values of which differ from the \grl\ result.

Further studies of the massless sector might help us to clarify the role
of torsion in the AdS/CFT correspondence.

\section*{Acknowledgements}

This work was supported by the Serbian Science Foundation under Grant No.
171031.

\appendix
\section{AdS and Siklos spacetimes in 3D}
\setcounter{equation}{0}

In this Appendix, we review basic aspects of the three-dimensional AdS and
Siklos spacetimes; see for instance \cite{4,31,32} and
\cite{7,8,11}, respectively.

The AdS space in 3D, with topology $S^1\times R^2$,  can be defined in
terms of the hypersurface
$$
H_3:\qquad \bu^2-\bx^2-\by^2+\bv^2=\ell^2\, ,
$$
embedded in a 4-dimensional Minkowski space with metric
$\eta_{ab}=(1,-1,-1,1)$. The metric on $H_3$ has the form
\be
ds^2=d\bu^2-d\bx^2-d\by^2+d\bv^2\, ,
\ee
its isometry group is $SO(2,2)$, and the scalar curvature is $R=6/\ell^2$.

The space $H_3$ can be covered by the global coordinates $(t,\r,\vphi)$,
\bea
&&\bu=\ell \cosh\r\cos t\, ,\qquad \bx=\ell\sinh\r\cos\vphi\,,  \nn\\
&&\bv=\ell\cosh\r\sin t\, ,\qquad  \by=\ell\sinh\r\sin\vphi\,,  \nn
\eea
with $t\in[-\pi,\pi], \r\in[0,\infty)$, for which the metric takes the
form
\be
ds^2=\ell^2\left[dt^2\cosh^2\r
                 -(d\r^2+\sinh^2\r d\vphi^2)\right]\, .
\ee
However, since $t$ is an angle, there are closed timelike curves in $H_3$.
The problem can be cured by replacing the $S^1$ time $t\in[-\pi,\pi]$ by a
new, $R^1$ time $t\in(-\infty,+\infty)$, changing thereby the topology
from $S^1\times R^2$ to $R^3$. The space obtained in this way is known as
the \emph{universal covering} of the AdS space. According to the commonly
accepted terminology, it is this space that is called the AdS space; we
denote it by \ads3. A simple form of the \ads3\ metric is obtained in the
Schwarzschild-like coordinates $r=\ell\sinh\r, \ell t\to t$.

Let us now parametrize \ads3\ by introducing the Poincar\'e coordinates:
$$
\t=\frac{-\bv}{\bu+\bx}\, ,\qquad
x=\frac{\by}{\bu+\bx}\, , \qquad y=\frac{\ell}{\bu+\bx}\, .
$$
They do not cover the whole space, but only one of the regions where
$\bu+\bx$ has a definite sign. In these regions, the metric has the form
\be
ds^2=\frac{\ell^2}{y^2}\left(2dudv-dy^2\right)\, ,              \lab{A.3}
\ee
where $u=(\t+x)/\sqrt{2}$, $v=(\t-x)/\sqrt{2}$, and the boundary is
located at $y=0$.

The Killing vectors $\xi=\xi^\m\pd_\m$ for the metric \eq{A.3} are defined
by the conditions
$$
\d_0 g_{\m\n}:=-\pd_\m\xi^\r g_{\r\n}-\pd_\n\xi^\r g_{\r\m}
               -\xi^\r\pd_\r g_{\m\n}=0\, .
$$
They produce a set of requirements on $\xi^\m$, the solutions of which
define a basis of six independent AdS Killing vectors $\xi_{(m)}$:
\bea
&& \xi_{(1)}=(\ell,0,0)\, ,\qquad  \xi_{(4)}=(0,2v,y)\, ,       \nn\\
&& \xi_{(2)}=(0,\ell,0)\, ,\qquad
   \xi_{(5)}=\left(\frac{u^2}{\ell},\frac{y^2}{2\ell},
                                    \frac{uy}{\ell}\right)\,,   \nn\\
&& \xi_{(3)}=(u,-v,0)\, ,\qquad
   \xi_{(6)}=\left(\frac{y^2}{2\ell},\frac{v^2}{\ell},
                            \frac{vy}{\ell}\right)\, .          \lab{A.4}
\eea

Turning now to the class of Siklos spacetimes \eq{2.1}, we note that it is
equivalent to a subclass of Kundt spacetimes, defined by the metric
\be
ds^2=2\left(\frac{q}{p}\right)^2dU(\bH dU+dV)-\frac{1}{p^2}dY^2\, ,\lab{A.5}
\ee
where $\bH=\bH(U,Y)$, and
$$
p:=1+\frac{\l}{4}Y^2\,,\qquad q:=\left(1+\sqrt{-\frac{\l}{4}}\,Y\right)^2\, ,
$$
with $\l:=-1/\ell^2$. Indeed, by introducing the new coordinates
$$
Y=-2\ell\frac{y+1/2}{y-1/2}\, ,\qquad
U=2\ell u\, , \qquad V=2\ell v\, ,
$$
one ends up with the Siklos metric \eq{2.1}, where the new function
$H=H(u,y)$ is defined by $H(u,y):=\bH(U,Y)|_{U=U(u),Y=Y(y)}$.

For general $H$, the only Killing vector of the Siklos metric is
$\xi_{(2)}=\ell\pd_v$, but for some specific forms of $H$ there can be
more Killing vectors; for instance, $\xi_{(1)}=\ell\pd_u$ when $H$ is
independent of $u$, or the maximal number of six Killing vectors \eq{A.4}
when $H=0$.
\section{PGT field equations}
\setcounter{equation}{0}

In this Appendix, we give a brief derivation of the PGT field equations,
based on Ref. \cite{22}. By varying the gravitational Lagrangian \eq{3.4}
with respect to $b^i$ and $\om^{ij}$, one obtains the PGT field equations
in vacuum, written in a compact form as
\bea
\first\quad &&\nab H_i+E_i=0\,,                                 \nn\\
\second\quad &&\nab H_{ij}+E_{ij}=0\, .                         \lab{B.1}
\eea
Here, $H_i:=\pd L_G/\pd T^i$ and $H_{ij}:=\pd L_G/\pd R^{ij}$ are the
covariant momenta, and $E_i:=\pd L_G/\pd b^i$ and $E_{ij}:=\pd
L_G/\pd\om^{ij}$ are the gravitational energy-momentum and spin currents,
respectively. In more details, we have
\bea
&&H_i=2a_1\hd {}^{(1)}T_i\, ,                                           \nn\\
&&H_{ij}=-2a_0\ve_{ijm}b^m
        +2\,\hd\left(b_4{}^{(4)}R_{ij}+b_6{}^{(6)}R_{ij}\right)\,,\lab{B.2}
\eea
and
\bea
&&E_i=h_i\hook L_G-(h_i\hook T^m)H_m-\frac{1}{2}(h_i\hook R^{mn})H_{mn}\,,\nn\\
&&E_{ij}=-(b_iH_j-b_jH_i)\, .
\eea
Now, substituting these expressions into \eq{B.1}, one obtains an explicit
form of the PGT field equations, displayed in Eqs. \eq{3.5} of the main
text.
\section{Improving the canonical generator}
\setcounter{equation}{0}

In this appendix, we construct the improved gauge generator for the
massless sector of our solution.

Gauge symmetries of the first-order Lagrangian \eq{5.1} are described by
the canonical gauge generator $G$, the form of which can be found in
Eqs. (5.7) of Ref. \cite{22}. To examine the differentiability of $G$, we
start from the form of its variation:
\bea
&&\d G=-\int_\S d^2x(\d G_1+\d G_2)\, ,                         \nn\\
&&\d G_1=-\ve^{t\a\b}\xi^\mu\left(b^i{_\mu}\pd_\a\d\t_{i\b}
         +\om^i{_\mu}\pd_\a\d\r_{i\b}+\t^i{_\mu}\pd_\a\d b_{i\b}
         +\r^i{_\mu}\pd_\a\d\om_i{_\b}\right)+R\, ,             \nn\\
&&\d G_2=-\ve^{t\a\b}\th^i\pd_\a\d\r_{i\b}+R\, .                \lab{B.1}
\eea
Here, the variation is performed in the set of asymptotic states, $R$
stands for regular terms and $\r^i$ is the Lie dual of $\r_{mn}$:
$$
\r_i=2\left(a_0-\frac{b_6}{6}R\right)b_i+
      2b_4\left(\ric_{(ik)}-\frac{1}{3}R\eta_{ik}\right)b^k\,.
$$
Moreover, the coherently oriented volume 2-form on $\S$, expressed in the
new coordinates $(t,\vphi,y)$, is normalized to $d^2x=dy d\vphi$. Together
with $\ve^{y\vphi}:=\ve^{ty\vphi}=1$, this is in accordance with the
conventions used in Ref. \cite{22}.

As one can see, $G$ is not differentiable, but the problem can be
corrected by going over to the improved canonical generator $\tG:=G+\G$,
where the surface term $\G$ is constructed so that $\d \tG=R$. In the
process, transition to surface integrals is performed with the help of the
Stokes formula:
$$
\int_\S d^2x\,\pd_\a v^\a=\int_{\pd\S}df_\a v^\a
              =\int_0^{2\pi} d\vphi v^y\,,\qquad df_\a=\ve_{\a\b}dx^\b\, .
$$
Thus, using \eq{B.1} and the asymptotic conditions \eq{4.5} and \eq{4.8},
the surface term $\G$ in the improved generator $\tG\equiv G+\G$ is found to
have the following form:
\bsubeq\lab{B.2}
\bea
\G&=&\G_u+\G_v\, ,\nn\\
\frac{\sqrt 2}{\ell}\G_u&=&-2\left(a_0-\frac{b_6}{\ell^2}\right)
   \int_0^{2\pi} d\vphi\eps^u\frac{1}{y}\left(B^-{_u}-B^-{_v}\right)
      +2\ell a_1\int_0^{2\pi} d\vphi
    \eps^u\pd_y \left(B^-{_u}-B^-{_v}\right)                    \nn\\
&&+\frac{2b_4}\ell\int_0^{2\pi} d\vphi\eps^u
        \left(\pd_y\Om^-{_u}-\pd_u\Om^-{_y}
              +\frac{1}{y}\frac{B^-{_u}}{\ell}\right)\, ,       \\
\frac{\sqrt 2}\ell\G_v&=&2\left(a_0-\frac{b_6}{\ell^2}\right)
    \int_0^{2\pi} d\vphi\eps^v\frac{\ell}{y}\left(\Om^+{_u}-\Om^+{_v}
    +\frac 1{\ell}B^+{_u}-\frac 1{\ell}B^+{_v}\right) \, .
\eea
\esubeq
The result for $\G_u$ is simplified with the help of the condition
$a_0-b_6/\ell^2-a_1=0$, which is used in Eq. \eq{3.9} to define the
massless sector of the torsion wave. The factors $\sqrt{2}/\ell$ appear as
an effect of the change of coordinates $(t,\vphi)\to(u,v)$ in the
components of $B^i$ and $\Om^i$.

The above construction shows that $\tG$ is differentiable provided it is
finite, and the finiteness of $\tG$ follows from the finiteness of
$\G\equiv\G_u+\G_v$. The term $\G_v$ is seen to be finite directly from the
adopted asymptotic conditions, whereas the finiteness of $\G_u$ depends on
the validity of an additional relation:
\be
-\left(a_0-\frac{b_6}{\ell^2}-\frac{b_4}{\ell^2}\right) B^-{_u}
         +\frac{b_4}{\ell}y\pd_y\Om^-{_u}=\cO_1\, .             \lab{B.3}
\ee
To clarify this situation, we note that the original set of the asymptotic
conditions, given in Eqs. \eq{4.5} and \eq{4.8}, can be extended using the
following general principle: the expressions that vanish on-shell should
have an arbitrarily fast asymptotic decrease, as no solution of the field
equations is thereby lost. This principle allow us to derive the needed
relation \eq{B.3} as the ($\mu=v$, $i=+$) component of the field equation
\be
\ve^{\mu\nu\r}\left(\nab_\mu\r_{i\nu}
           +\ve_{ijk}b^j{_\n}\t^k{_\r}\right)=0\, .             \lab{B.4}
\ee

The surface terms \eq{B.2} are used in section \ref{sec5} to calculate the
canonical algebra of the improved gauge generators.  Note, in particular,
that $\G$ vanishes on the AdS background.


\end{document}